\documentclass[aps,prl,twocolumn,showpacs,letterpaper,superscriptaddress]{revtex4-1}

\usepackage{graphicx,dcolumn,longtable,epsfig}
\usepackage[usenames]{color}
\usepackage{amssymb}
\usepackage{amsmath}
\usepackage{epstopdf}
\usepackage{bm}
\usepackage{footnote}

\input epsf.tex

\def\jpb#1#2#3{J.~Phys.~B:~{\bf #1},\ #2\ (#3)}

\def\pra#1#2#3{Phys.~Rev.~A~{\bf #1},\ #2\ (#3)}
\def\prb#1#2#3{Phys.~Rev.~B~{\bf #1},\ #2\ (#3)}

\def\rmp#1#2#3{Rev.~Mod.~Phys.~{\bf #1},\ #2\ (#3)}
\def\prl#1#2#3{Phys.~Rev.~Lett.~{\bf #1},\ #2\ (#3)}

\def\njp#1#2#3{New~J.~Phys.~{\bf #1},\ #2\ (#3)}

\def\rmp#1#2#3{Rev.~Mod.~Phys.~{\bf #1},\ #2\ (#3)}
\def\jphc#1#2#3{J.~Phys.~C.~{\bf #1},\ #2\ (#3)}

\def\rpp#1#2#3{Rep.~Prog.~Phys.~{\bf #1},\ #2\ (#3)}

\def\etal{{\it et al.}}
\def\bea{\begin{eqnarray}}
\def\eea{\end{eqnarray}}
\def\be{\begin{equation}}
\def\ee{\end{equation}}

\usepackage{color}

\begin{document}
\title{Quantum Phases of Dipolar Bosons in Bilayer Geometry}
\author{ A.~Safavi-Naini}
\affiliation{Department of Physics, Massachusetts Institute of Technology, Cambridge, Massachusetts, 02139, USA}
\affiliation{ITAMP, Harvard-Smithsonian Center for Astrophysics, Cambridge, Massachusetts, 02138, USA}
\author{\c{S}.~G.~S\"{o}yler}
\affiliation{The Abdus Salam International Centre for Theoretical Physics, Strada Costiera 11, I-34151 Trieste, Italy}
\author{ G.~Pupillo }
\affiliation{ISIS (UMR 7006) and IPCMS (UMR 7504), Universit\'e de Strasbourg and CNRS,
Strasbourg, France}
\author{H.~R.~Sadeghpour}
\affiliation{ITAMP, Harvard-Smithsonian Center for Astrophysics, Cambridge, Massachusetts, 02138, USA}
\author{ B.~Capogrosso-Sansone }
\affiliation{Homer L. Dodge Department of Physics and Astronomy, The University of Oklahoma, Norman, Oklahoma 73019, USA }

\begin{abstract}
We investigate the quantum phases of hard-core dipolar bosons confined to a square lattice in a bilayer geometry. Using exact theoretical techniques, we discuss the many-body effects resulting from pairing of particles across layers at finite density, including a novel pair supersolid phase, superfluid and solid phases. These results are of direct relevance to experiments with polar molecules and atoms with large magnetic dipole moments trapped in optical lattices.
\end{abstract}

\maketitle

Recent experimental breakthroughs in the realization of ultracold gases of high-spin aligned atoms with large dipole moments~\cite{Pfau,Laburthe-Tolra,Lev,Ferlaino}, highly excited Rydberg atoms~\cite{RydbergReviews,Comparat}, and  of ground-state polar molecules~\cite{JILA,Naegerl} hold considerable promise for  investigations of many-body quantum systems where dipolar interactions can become dominant~\cite{PfauReview,CarrReview,BarbaraReview,Baranov2012}. The anisotropy of dipolar interactions combined with the possibility to confine particles in low dimensional geometries using optical lattices allow for study of novel  pairing mechanisms and the associated quantum phases in a setup where collisional losses are suppressed. This is particularly intriguing for the case of magnetic atoms, where confinement to lattices with spacings as small as 200nm is possible~\cite{Lev_PrivateComm}, which favors inter-site dipolar interactions and pairing.

Pairing of two spin-polarized {\it fermionic} dipoles across coupled two-dimensional (2D) layers~\cite{wang2006}
or one-dimensional (1D) wires~\cite{Kollath} in an optical lattice has already lead to the
prediction of 2D inter-layer superfluidity~\cite{shlyapnikov2010,baranov2010,potter2010,ZinnerAll,Bruun,MarchettiParish2012}, analogous to bi-exciton condensation, and the 1D quantum
roughening transition~\cite{kuklov2011} in the case of equal number of particles in each layers. Additional exotic phenomena occur for unequal populations~\cite{worksDemler}, where (spin-rotational) symmetry breaking can induce, e.g., stable liquids and crystals of composite multimers~\cite{Dalmonte2011}.
For {\it bosonic} gases in the strongly interacting regime~\cite{Santos}, emergent parafermionic behavior has been demonstrated~\cite{Lecheminant2012,Kuklov2012} in coupled 1D wires. In two dimensions, a recent mean-field study in bilayer geometry~\cite{Trefzger2009}
has predicted novel quantum phenomena for a model of dipolar bosons on a lattice, including a so-called pair supersolid phase. Different from supersolids on a single lattice~\cite{ProkofevBoninsegniRMP2012, Capogrosso2010, PolletPRL2010}, the latter implies diagonal solid order coexisting with an off-diagonal superfluid order, both derived from composite pairs of dipoles. The experimental observation of this quantum phase and the associated pair superfluids and solids would constitute a breakthrough for condensed matter in the cold atomic and molecular context. Thus, the challenge is now to determine whether these quantum phases can be realized for realistic Hamiltonian representing the microscopic dynamics of strongly interacting dipolar bosons as realized in experiments.

In the present work we study a system consisting of hardcore dipolar bosons confined to two neighboring two-dimensional (2D) layers of a 1D optical lattice. The dipole moment of each particle is polarized perpendicular to the layers, which results in repulsive in-plane dipole-dipole interactions. This ensures collisional stability against short-range inelastic collisions in the strongly interacting gas. Out-of-plane dipolar interactions are dominantly attractive, which favors inter-layer pairing. Using exact theoretical techniques based on quantum Monte-Carlo methods~\cite{Worm}, below we demonstrate that this anisotropy and the long-range nature of interactions can induce crystallization of the dipolar cloud into a charge-density wave for a wide range of trapping parameters and interactions. Exotic quantum phases such as the pair-supersolid (PSS) phase and a pair-superfluid (PSF) are achieved under experimentally realistic trapping conditions. These phases can survive up to temperatures of the order of a few nK for a gas of polar molecules or strongly magnetic atoms.

\begin{figure}[b]
\begin{center}
\includegraphics[width=0.45\textwidth]{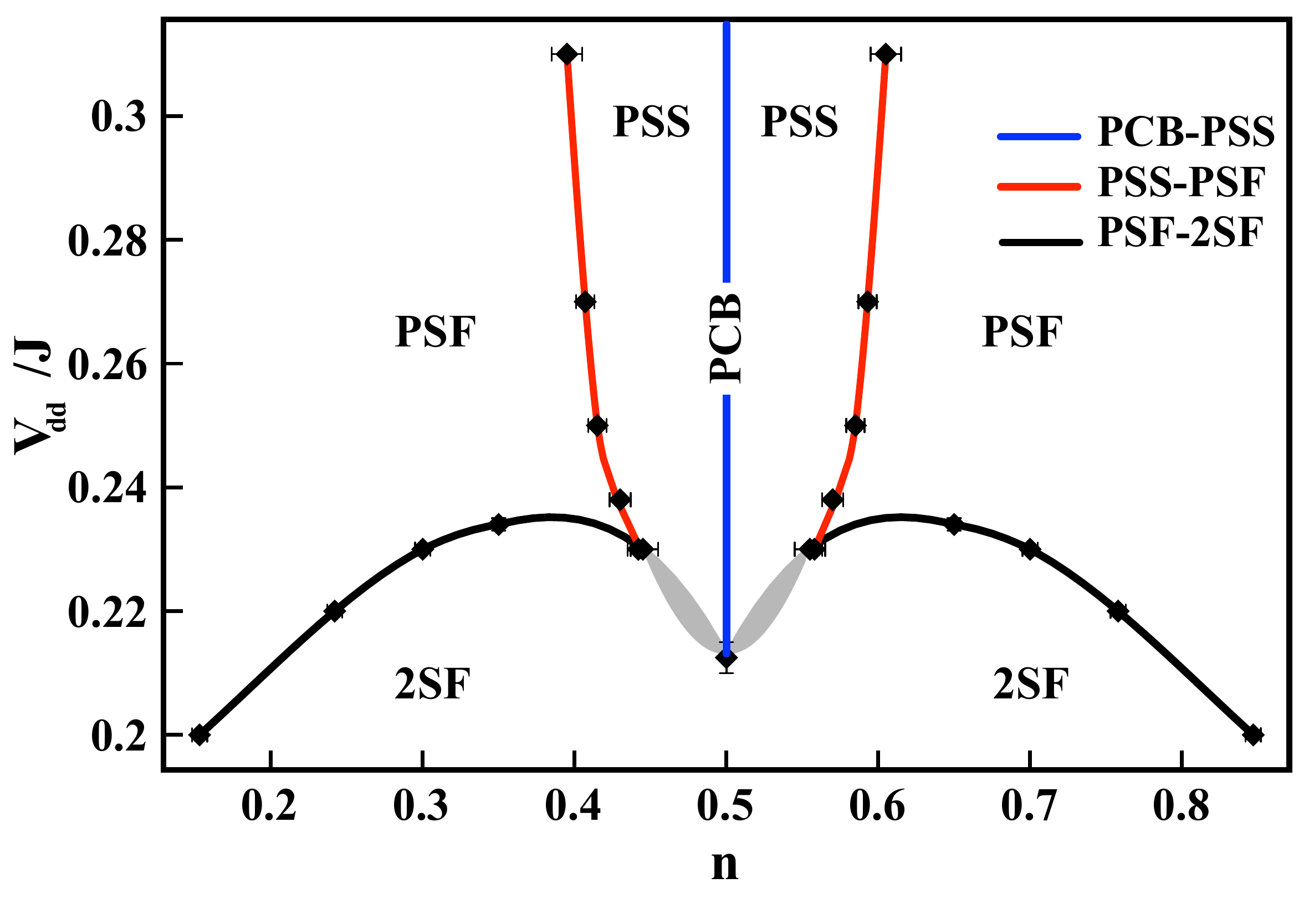}
\caption{(Color online) Phase diagram of Hamiltonian~\eqref{eq:H} as a function of $V_{\text dd}/J$ and particle density $n$, computed via QMC simulations, for an interlayer distance $d_z/a = 0.36$ (see text). CB: Checkerboard solid; PSS: Pair supersolid; PSF: Paired superfluid; 2SF: independent superfluids. The phase boundaries in the dashed region are not resolved.}
\label{fig:PD}
\end{center}
\end{figure}
The system we have in mind is described by the single-band tight-binding Hamiltonian
\begin{align}
\label{eq:H}
\nonumber H= &-J\sum_{<i,j>,\alpha}a_{i\alpha}^\dagger a_{j\alpha}-\frac{1}{2}\sum_{i\alpha;j\beta}V_{i\alpha;j\beta}n_{i,\alpha}n_{j,\beta}\\
&-\sum_{i,\alpha}\mu_{\alpha} \; n_{i,\alpha}.
\end{align}
Here $\alpha, \beta = 1,2$ and $i,j$ label the layers and the lattice sites in each layer, respectively, while $a_{i,\alpha}$ ($a_{i,\alpha}^\dagger$) are the bosonic creation (annihilation) operators, with $a_{i,\alpha}^{\dagger\,2}=0$ and $n_{i,\alpha}=a_{i,\alpha}^\dagger a_{i,\alpha}$. The brackets $<>$ denote summation over nearest neighbors only. The first term in Eq.~\eqref{eq:H} describes the kinetic energy with in-plane hopping rate $J$. The second term is the dipole-dipole interaction given by  $V_{i\alpha;j\beta}=C_{\text{dd}}(1-3\cos^2\theta)/(4\pi \vert i_{\alpha}-j_{\beta}\vert^3)$, where $\theta$ is the angle between particles at positions $i_{\alpha}$ and $j_{\beta}$ and $C_{\text{dd}}=d^2/\epsilon_0\,(C_{\text{dd}}=\mu_0d^2)$ for electric (magnetic) dipoles of strength $d$.
We denote the repulsive (attractive) nearest neighbor intra-layer (inter-layer) interaction by $V_{\text{dd}}= C_{\text{dd}}/(4\pi a^3)$ ($V_{\text{dd}}^\perp=2 C_{\text{dd}}/4\pi d_z^3$), with $a$ the in-plane lattice constant. The interlayer distance is $d_z$. The relative strength $V_{\text{dd}}/V_{\text{dd}}^\perp$ can be tuned over a wide range of values by changing $d_z/a$. The quantity $\mu_{\alpha}$ is the chemical potential which sets the number of particles in each layer. Here we fix $\mu_1=\mu_2$, i.e. $N_1=N_2$

Hamiltonian~\eqref{eq:H} provides a microscopic description for the dynamics of, e.g., a gas of RbCs molecules ($d\approx 1.25$~Debye) at low-density $n$, such that the initial system has no doubly occupied sites~\cite{BuchlerNat2007}. Collisional stability is ensured for $n^{-1/2} \gg (d^2/\hbar \omega_\perp)^{1/3} \simeq 130$nm with $\omega_\perp \simeq 100$kHz the frequency of transverse confinement provided by the in-plane optical lattice~\cite{Buchler2007}. In addition, the choice $d^2/d_z^3 < V_0$ avoids interaction-induced inter-layer tunneling, with $V_0$ the depth of the optical potential in the transverse direction. Model~\eqref{eq:H} can also be used to describe the dynamics of a gas of strongly magnetic dipolar atoms, such as Dy ($d=10\mu_B$). In this case the conservative estimate above for collisional stability is satisfied for $\omega_\perp \simeq $~1 kHz.

In the following, we present exact theoretical results based on path integral Quantum Monte Carlo simulations using a two-worm algorithm \cite{Worm2} which allows for efficient sampling of paired phases. We have performed simulations of $L\times L=N_{sites}$ square lattices with $L=8, 12, 16, 20$ and $24$. For computational convenience, we have set the dipole-dipole interaction cutoff to the third nearest neighbor and have checked that using a larger cutoff did not change the simulation results within errorbars. Lower cutoff values do not allow for stabilization of, e.g., supersolid phases, see below. In the following we choose $d_z/a=0.36$ and $d_z\sim 200$~nm, which is experimentally feasible with, e.g., Cr or Dy atoms~\cite{Laburthe-Tolra, Lev}. We show below that this choice allows one to access a parameter regime where particles on different layers can pair up to form a composite object. Below, we first discuss the phase diagram, and then discuss in more details the various phases.\\

\begin{table}
\begin{center}
    \begin{tabular*}{0.20\textwidth}{ c|c|c|c }

    \textbf{Phase} &$ S(\pi,\pi)$&$\psi_i^\alpha$ & $\Psi$ \\ \hline
        \hline
    PCB& $\neq0$& 0 & 0 \\ \hline
    PSS & $\neq0$& 0 & $\neq0$ \\ \hline
    PSF & 0& 0 & $\neq0$ \\ \hline
    2SF & 0& $\neq0$ & $\neq0$ \\ \hline
        \end{tabular*}
    \caption{Quantum phases of Fig.~\ref{fig:PD} and corresponding order parameters: structure factor $S(\pi,\pi)$; single-particle condensate $\psi_i^\alpha=\langle a_{i,\alpha}\rangle$ in each layer $\alpha$; pair-condensate order parameter $\Psi=\langle a_{i,\alpha}a_{i,\beta}\rangle$, with $\alpha \neq \beta$.  (See text)}
    \label{table:conditions}
\end{center}
\end{table}

{\it The phase diagram} of Eq.~\eqref{eq:H} at temperature $T=0$ is shown in Fig.~\ref{fig:PD} as a function of $V_{\text dd}/J$ and the density $n$, in the parameter regime $0.31 > V_{\text dd}/J > 0.2$ and $0.1 < n < 0.9$, with $d_z/a = 0.36$. We expect this phase diagram to be representative of situations with $d_z/a \ll 2$, where interlayer pairing is favored (see figure~\ref{fig:vvsd} below).\\ 

At half-filling $n=0.5$,  an incompressible checkerboard solid of pairs (PCB) is stabilized for sufficiently large values of $V_{\text{dd}}/J$. Similar to the conventional checkerboard phase present in single-layers \cite{Capogrosso2010}, here atoms in each layer occupy every other site of the lattice, due to in-plane dipolar repulsion. The checkerboard order is characterized by a finite value of the static structure factor $S(\mathbf{k})$ at the reciprocal lattice vector $\mathbf {k}=(\pi, \pi)$, with\begin{equation}
S(\mathbf{k})=\frac{1}{N}\sum_{r, r\prime} \exp[i\mathbf{k}(\mathbf{r}-\mathbf{r}^\prime)]\langle n_r n_{r^\prime} \rangle,
\end{equation} and the system displays zero superfluidity.
We find that in the PCB phase atoms across the layers are strongly paired due to attractive interlayer interactions. As a result, the position of the two checkerboard solids is strongly correlated, i.e., they sit on top of each other. The system can be thus envisioned as a solid of pairs~\cite{Trefzger2009, Capogrosso2011}, with an effective mass $m_{\rm{eff}}\sim J^2/(2V_{\rm{dd}}^\perp+zV_{\rm{dd}})$, where $z$ is the coordination number. The PCB solid is stabilized at (much) lower values of $V_{{\rm dd}}/J$ compared to the case of checkerboard solids in a single layer~\cite{Capogrosso2010}, in analogy with what found in~\cite{Capogrosso2011}. This is due to the higher effective mass of the pairs. A similar robustness of this phase is also found for melting at finite temperature.

Upon doping the PCB solid with extra particles or holes, a so-called pair-supersolid (PSS) phase is immediately stabilized. The latter displays both diagonal long range order with  $S(\pi,\pi)\ne0$, off-diagonal long-range order associated with a non-vanishing value of the pair-condensate order parameter $\Psi=\langle a_{i,\alpha}a_{i,\beta}\rangle\neq0$ (with $\alpha \neq \beta$), and an associated finite superfluid stiffness for pairs (see below). The single-particle condensate order parameter $\psi_i^\alpha=\langle a_{i,\alpha}\rangle=0$ is instead zero. The existence of off-diagonal order is consistent with a picture of delocalized defects~\cite{Andreev,ProkofevBoninsegniRMP2012}, which here correspond to correlated pairs of holes or extra particles across the layers. The PSS phase forms a lobe structure in the ($V_{\text{dd}}/J - n$)-plane, around the PCB line. Away from the tip of the lobe, we find that by varying $n$ at constant $V_{\text{dd}}/J$ the PSS loses its diagonal long range order by melting into a pair superfluid phase (PSF), via an Ising-type transition (red continuous line).  The PSF phase, with $\Psi\neq0$ and $\psi_i^\alpha=0$, is destroyed in favor of   a 2SF (a phase with independent, though correlated, superfluids on each layer) for smaller values of $V_{\text{dd}}/J$. In particular, we notice that a tiny PSF-region should persist in between the PSS and 2SF phases even close to the tip of the PSS-lobe, however this is within errorbars for $V_{\text{dd}}/J \lesssim 0.2$. Exactly at filling $n=0.5$, our results are consistent with a direct PCB-2SF transition, as discussed below, with no intermediate PSS phase. In particular we find no evidence of, e.g., possible micro-emulsion phases~\cite{Spivak,PolletPRL2010}, within errorbars.

Finally, we notice that a host of other phases are present in the general phase diagram for two layers. In particular, we find that for stronger values of $V_{\text{dd}}/J\gtrsim 0.3$  the system displays a sequence of incompressible phases at various rational fillings of the lattice, similar to the so-called Devil's staircase found in the case of a single layer. We also expect novel PSS phases to appear around lobes at, e.g., filling $n=0.25$, in analogy with Ref.~\cite{Capogrosso2010}. In addition, independent solids as well as supersolid phases can be achieved by increasing the layer distance, while mixtures of solid and superfluid phases can be stabilized by modifying the relative particle density in the two layers. The discussion of some of these phases is however outside of the scope of the present work. In the remainder of the paper we discuss in more detail the various phases and their transitions at zero and finite temperature around $n=0.5$.\\

{\it Stability of the PCB phase:} As discussed above, the PCB phase at $n=0.5$ is characterized by a finite value of the order parameter $S(\pi,\pi)$ and no off-diagonal order. The latter is associated with superfluidity in a (2+1) dimensional interacting system, which can be measured straightforwardly within Monte-Carlo (see below). In addition, within the PCB phase inter-layer dipolar attraction strongly correlates the positions of particles in the two layers.

The stability of the PCB phase with respect to intra-plane interactions as well as inter-layer distance $d_z/a$ at zero temperature is analyzed in Fig.~\ref{fig:vvsd}. There, we numerically determine the minimum dipolar interaction strength $V_{\text dd}/J$ required to stabilize the PCB phase at a given $d_z/a$. In order to establish whether the solid phase is paired we have performed several simulations with different initial conditions for each set of parameters and observed whether the equilibrium configuration was dependent on the initial choice or not. The figure shows that a PCB phase is stabilized for $d_z/a \lesssim 2$ and sufficiently large $V_{\text dd}/J$ (continuous line). In this parameter regime, the system above (below) the continuous line is a PCB (2SF) phase, respectively, that is, the continuous line visualizes the shift of the PCB-2SF transition point of Fig.~\ref{fig:PD} as a function of $d_z/a$.
Instead, for $d_z/a > 2$ and large enough interactions the insulating phase above the (dotted) line corresponds to two independent checkerboard phases (2CB). This points to the possible presence of a tri-critical point in the phase diagram around $d_z/a \approx 2$. We have confirmed that the computed transition points are independent of the interaction cutoff that we use, within our errorbars, and should be thus quantitatively relevant to experiments.

In the following we focus on  $d_z/a = 0.36$ to satisfy $V_{dd}^\perp\gtrsim 10J$ in the vicinity of the tip of the lobe. We find that this choice ensures pairing at $n\sim 0.5$ (in the vicinity of the PCB phase) while keeping $V_{dd}$ relatively low. This corresponds to experimentally optimal conditions to observe PSS phase: a lower effective mass of pairs $m_{\rm eff}$ results in a larger superfluid density which in turn results in higher critical temperatures (see also below).

\begin{figure}[h]
\begin{center}
\includegraphics[width=0.45\textwidth]{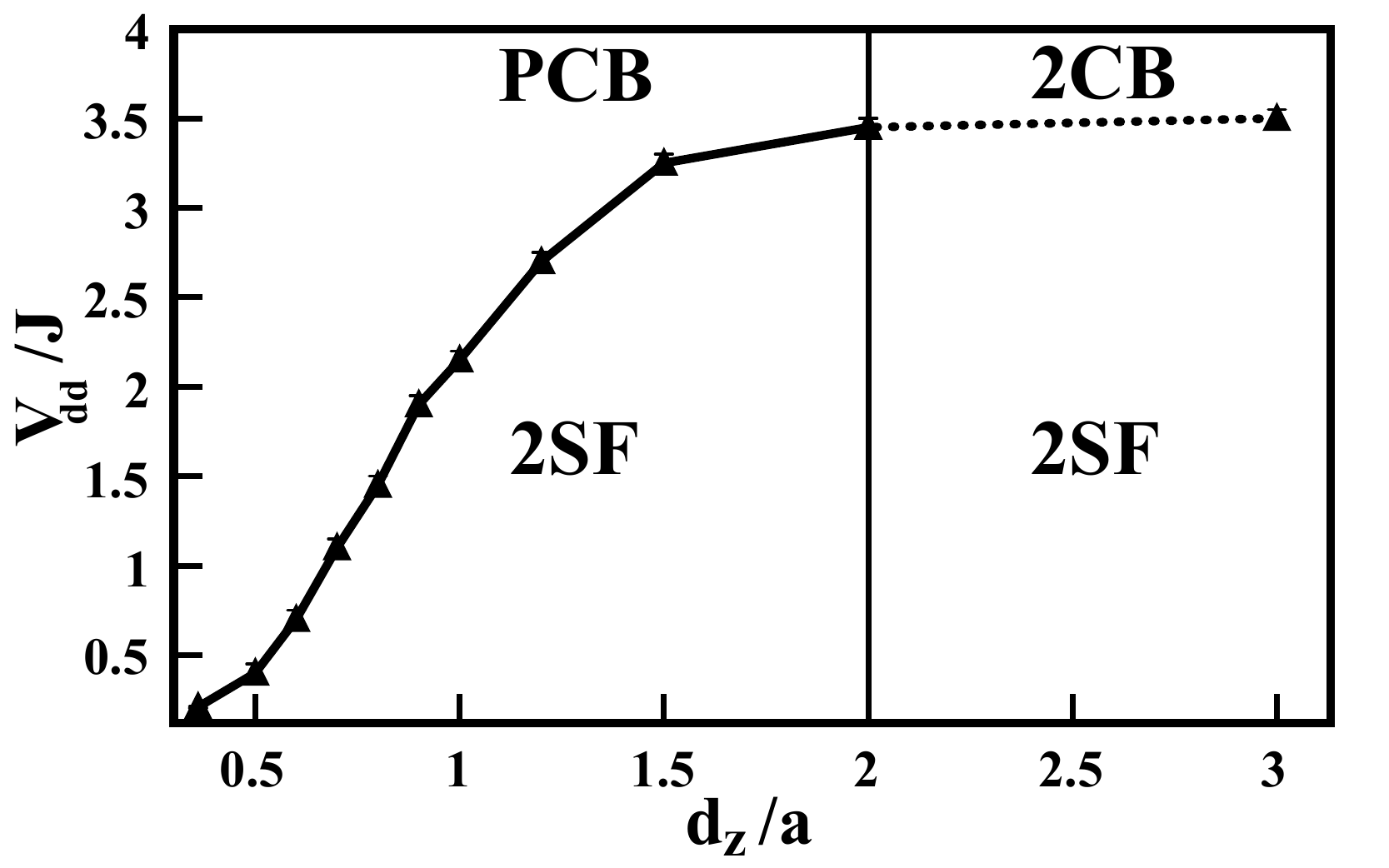}
\caption{(Color online) The plot of minimum $V_{dd}/J$ needed to stabilize the CB phase as a function of $d_z/a$. Once the layers are separated by $d_z/a>2$ they behave as independent layers. }
\label{fig:vvsd}
\end{center}
\end{figure}

{\it Pair supersolid phase:} Figure~\ref{fig:PD} shows that a PSS lobe is immediately formed by doping the PCB solid with either vacancies (holes) or interstitials (extra particles). The hard-core constraint of Eq.~\eqref{eq:H} ensures particle-hole symmetry, and thus reflection symmetry of the lobe, around $n=0.5$.

\begin{figure}[h]
\begin{center}
\includegraphics[width=0.45\textwidth]{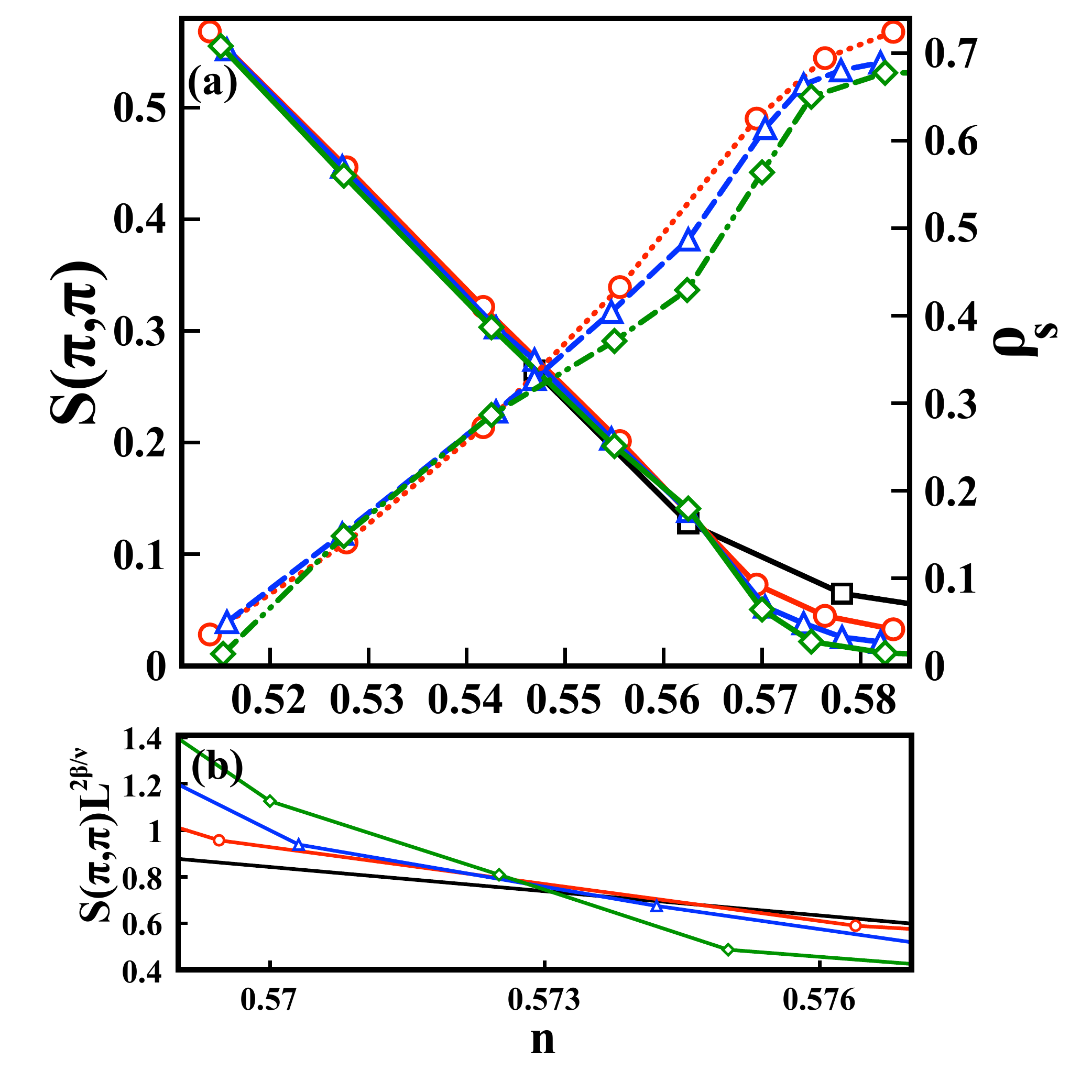}
\caption{(Color online) (a) Structure factor $S(\pi,\pi)$ (solid lines, left $y$-axis) and superfluid stiffness $\rho_s$ (dashed lines, right $y$-axis) in the PSS phase for $L=8$ (black squares), 12 (red circles), 16 (blue triangles) and 20 (green diamonds) at $T/J=1/(1.5 L)$ shown using black squares, red circles, blue triangles and green diamonds respectively. The PSS-PSF transition point is at $V_{\text dd}=0.238J$. (b) Scaled structure factor $S(\pi,\pi)L^{2\beta/\nu}$ vs. n with $2\beta/\nu=1.0366$ for $L=8, 12, 16$ and $20$.}
\label{fig:PSS}
\end{center}
\end{figure}

We characterize this pair supersolid phase in Fig.~\ref{fig:PSS}, for a specific choice of interaction strength $V_{dd}/J=0.238$. In the figure, the order parameter for the diagonal checkerboard solid order $S(\pi,\pi)$ (continuous lines) and the superfluid stiffness of {\it pairs} $\rho_{PSS}$ are plotted as a function of $n$. The quantity  $\rho_{PSS}=T\langle \mathbf{W}^2 \rangle /d L^{d-2}$ \cite{winding} is directly related to a pair condensate, and can be calculated within quantum Monte-Carlo, with $\mathbf{W}=W_1+W_2$ the sum of winding numbers in layer 1 and 2. The figure shows that for an extended range of densities, both the static structure factor and the pair superfluid stiffness are finite and system size independent, showing the existence of a stable supersolid phase in the lobe region. We note that, due to pairing across the layers, in the PSS phase the fluctuation of {\it difference} in winding numbers is zero $\langle (W_1 - W_2)^2\rangle$.

{\it Superfluid phases:} As the system is doped further, the PSS disappears in favor of a PSF phase. The latter displays pair-induced off-diagonal long range order, only [see, e.g., Table 1]. We find that the PSS-PSF transition is of the Ising type universality class in (2+1)-dimensions, analogous to the case of a single-layer~\cite{Capogrosso2010}. Critical points are determined using finite size scaling for the static structure factor with scaling coefficients $2\beta/\nu=1.0366$~\cite{Ising} (see Fig.~\ref{fig:PSS}(b) for the specific choice $V_{\text dd}=0.238J$). In the figure the scaled quantity $S(\pi,\pi) L^{1.0366}$ is plotted as a function of $n$, and the crossing of the curves at $n_{\rm cr}= 0.573 \pm 0.002$ corresponds to the quantum critical point where the finite size effects disappear [see also panel (a)].\\

We find in general that by lowering the interaction strength $V_{\rm dd}/J$ at constant $n$ from the PSF phase, the system finally develops into two independent superfluids ($2$SF) with a finite value of the single-component condensate order parameters, $\psi_i^\alpha=\psi_i^\beta \neq0$, via a second order phase transition in the (2+1) XY universality class. The transition points between the PSF and 2SF phases in Fig.~\ref{fig:PD} are calculated using finite size scaling of $\langle (W_1 - W_2)^2\rangle$. The latter quantity is zero inside the PSF phase in the thermodynamic limit due to pairing across the layers, while it has a finite value in the 2SF phase. We note that the pair order parameter in the 2SF phase is instead trivially non-zero, $\Psi\neq0$ (see also Table 1).

The phase diagram in Fig.~\ref{fig:PD} shows that the boundary of the PSF-2SF transition shifts downward approximately linearly in the ($V_{\rm dd}/J - n$)-plane, as the density becomes sufficiently smaller or larger than $n=0.5$. This is easily understood in the limit of very small densities, by noting that inter-plane dipole-dipole interactions always favor the existence of a two-body bound state, even for an arbitrarily small interaction strength. However, we find that many-body effects result in a threshold for the formation of pairs at finite density, where the magnitude of the interaction strength required to stabilize pairing increases with $n$. This is explained by noting that, in the limit of low density, $d_z \sqrt{n} \ll 1$, PSF phase is composed of weakly interacting superfluid dimers. As the density is increased exchanges between dimers are favored. This destabilizes the dimers, inducing the transition to two independent superfluids in the 2SF phase. Eventually, the presence of diagonal order near $n=0.5$ forces the PSF-2SF line to bend down, deviating from the linear dependence on $n$. \\

We gain further insight into the structure of correlations in the condensed phases by studying the following four-point correlation function:
\begin{eqnarray}
f_{jl}=\langle \psi_{1,i}\psi_{2,i}\psi_{1,j}^{\dagger}\psi_{2,l}^\dagger\rangle.\label{eq:FourCorr}
\end{eqnarray}
Here $i$,$j,l$ refer to sites, $1$, $2$ refer to layers, and $\langle \rangle$ denotes a quantum and thermal average as well as site averaging over $i$.  In the presence of pair superfluidity, one expects this correlation function to be {\it short ranged} with respect to $r_{jl}=\vert r_j-r_l \vert$, and simultaneously {\it long ranged} with respect to $r_{il}=\vert r_i-r_l \vert$ and $r_{ij}=\vert r_i-r_j \vert$. In the 2SF phase, instead, $f_{jl}$ is obviously long ranged with respect to $r_{il}$ and $r_{ij}$, but it is {\it independent} of $r_{jl}$.

\begin{figure}
\begin{center}
\includegraphics[width=0.75\columnwidth]{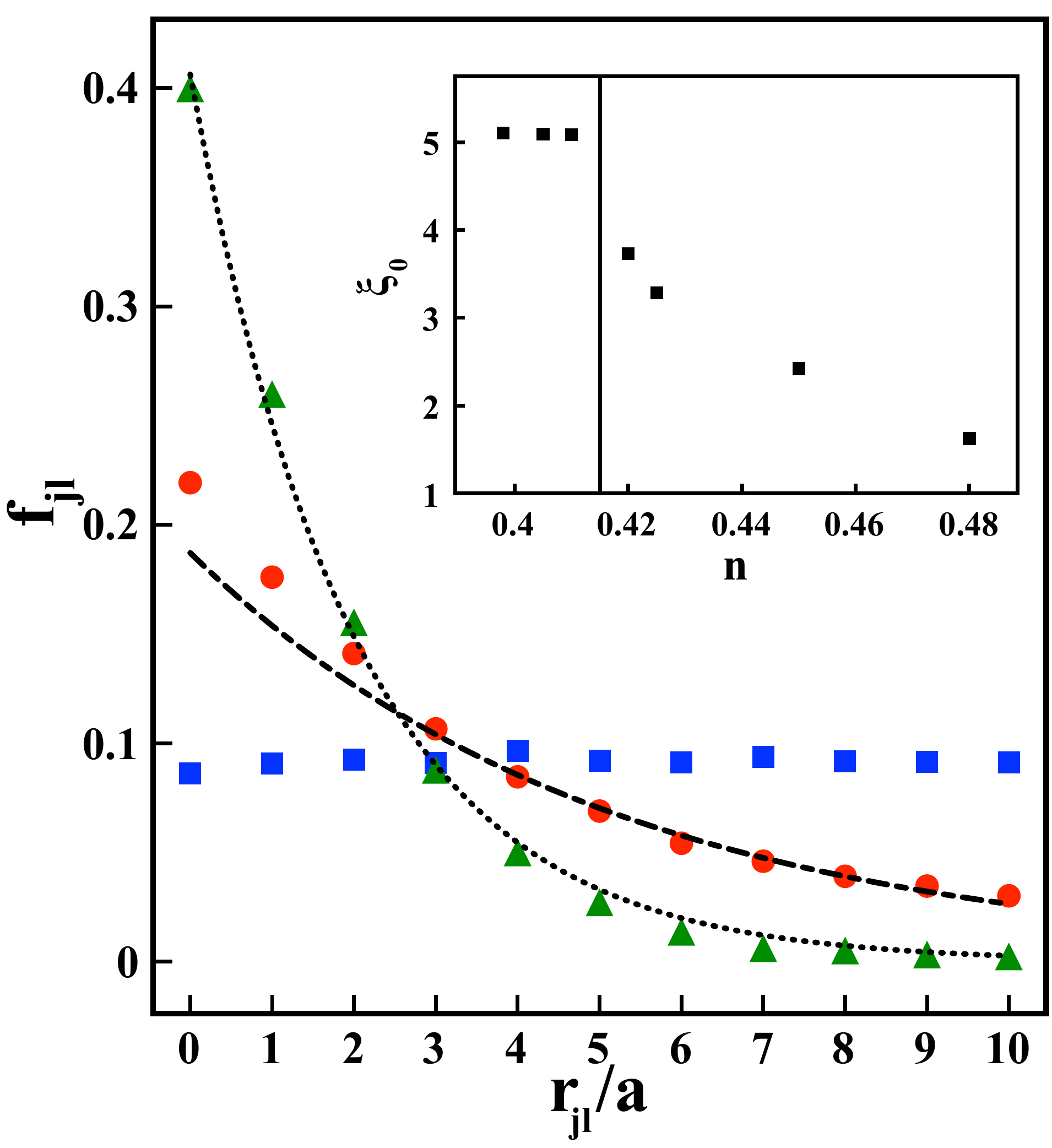}
\caption{(Color online) Four-point correlation function $f_{jl}$ of Eq.~\eqref{eq:FourCorr} as a function of $r_{jl}$ for the $2$SF (blue squares, $n=0.30,\,V_{\rm dd}=0.18J$), PSF  (red dots, $n=0.40,\,V_{\rm dd}=0.25J$) and PSS  (green triangles, $n=0.48, \,V_{\rm dd}=0.25J$) phases. The dashed (dotted) line is the exponential fit, $f_0e^{-r_{jl}/\xi_0}$, to the PSF (PSS) histogram, where $\xi_0$ can be interpreted as the extent of the pair wavefunction (see text). The inset shows $\xi_0$ across the PSF-PSS phase boundary.}
\label{fig:hist}
\end{center}
\end{figure}

Figure~\ref{fig:hist} shows $f_{jl}$ (normalized to unity: $\int_0^\infty f_{jl}dr_{jl}=1$) as a function of $r_{jl}$ for the PSS (green triangles, $n=0.48, \,V_{\rm dd}=0.25J$), PSF (red dots, $n=0.40,\,V_{\rm dd}=0.25J$) and 2SF phases (blue squares, $n=0.30,\,V_{\rm dd}=0.18J$). As expected, we find that $f_{jl}$ is independent of $r_{jl}$ in the 2SF phase, where pairing is absent, while it is peaked at $r_{ij}=0$ both in the PSS and PSF phase. The figure shows that an  exponential ansatz of the form $f_0 e^{-r_{jl}/\xi_0}$ fits quite well the large-$r_{ij}$ behavior of $f_{jl}$ in these latter phases, and is essentially exact for all $r_{jl}$ in the PSS phase with $\xi_0=1.63a$. Here $\xi_0$ can be interpreted as the spread of the pair wavefunction, and is obtained from Fig.~\ref{fig:hist} by fitting the tail of $f_{jl}$, as obtained numerically. The inset in Fig.~\ref{fig:hist} shows $\xi_0$ as a function of $n$, as the PSF-PSS phase boundary is crossed. The pair wavefunction is shown to be considerably more tightly bound in the PSS phase than in the PSF phase. The abrupt drop in $\xi_0$ locates precisely the transition point.\\

{\it Finite temperature:} We have studied the robustness of the quantum phases  described above against thermal fluctuations. As expected for two-dimensional systems, we find in general that superfluidity in the PSS, PSF and 2SF phases disappears at finite temperature $T$ via a Kosterlitz-Thouless (KT) type~\cite{KTtrans} transition. Diagonal long range order in the PCB and PSS phases is instead lost via a two-dimensional Ising-type transition. We have found that, when present, pairing still exists at the transition points, suggesting that the temperatures required for breaking pairs are higher than the critical temperatures measured here.

\begin{figure}[h]
\begin{center}
\includegraphics[width=0.45\textwidth]{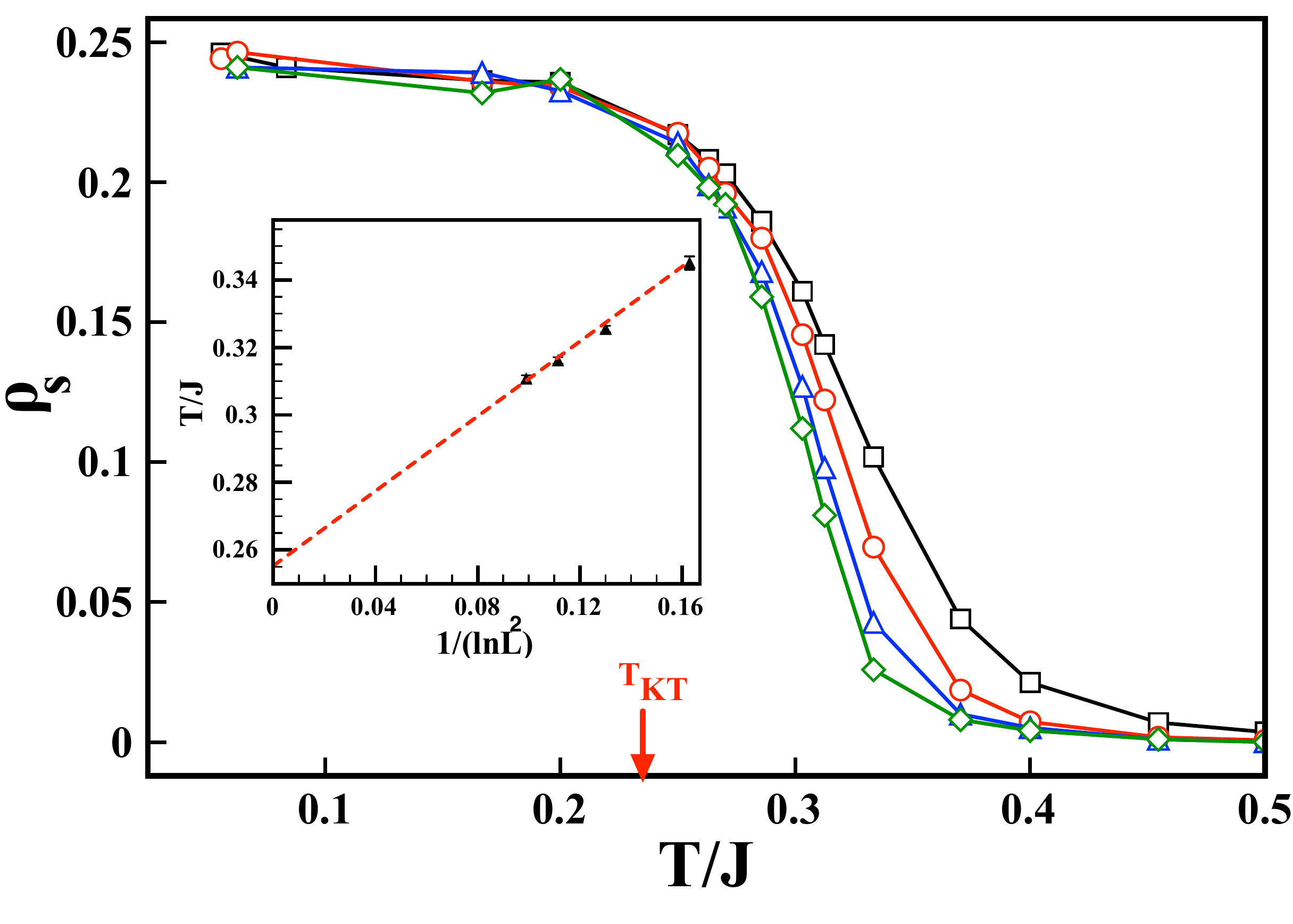}
\caption{(Color online) Superfluid stiffness $\rho_s$ as a function of temperature $T/J$, at $V_{\text dd}=0.20J$ and $n=0.30$, corresponding to $2$SF phase at $L=12, 16, 20$ and $24$ shown using black squares, red circles, blue triangles and green diamonds respectively. As temperature is increased the $2$SF phase undergoes KT phase transition at critical temperature $T_{KT,2SF}\approx 0.255J$, indicated by an arrow. The inset shows finite size scaling~\cite{Ceperly} where the dashed line is a linear fit of our simulation results (points). }
\label{fig:BKT}
\end{center}
\end{figure}

Figure \ref{fig:BKT} shows one example for the SF-normal transition in the 2SF phase. We plot $\rho_s$ vs. $T/J$ at $V_{\text dd}/J=0.20$ and $n=0.3$ for different system sizes. The inset shows the finite size scaling procedure~\cite{Ceperly} used to determine the critical temperature. We find $T_{KT,2SF}=\frac{\pi \hbar^2 \rho_s(T_{KT})}{2}\approx 0.255J$.
For the PSF-normal transition we find $T_{KT,PSF}\approx0.08 J$ at $n=0.3$ and $V_{\text dd}/J=0.25$. The lower KT transition temperature compared to the 2SF-normal transition is due to a larger effective mass of the pairs, i.e. lower effective hopping, which results in a suppression of particle delocalization and consequently smaller $\rho_s$.
The disappearance of the PSS phase proceeds in two successive stages. At $T_{KT, PSS}$ the PSS phase melts into a liquid-like phase reminiscent of a liquid crystal, with $\rho_s=0$ and $S(\pi,\pi)\neq 0$. Upon further increasing the temperature $S(\pi,\pi)$ becomes zero at a critical temperature $T_c$ through an Ising-type transition ($2\beta/\nu=1/4$ in 2D). For example, we find $T_{KT, PSS}\approx 0.06J$ and $T_c\approx 0.3J$ for $n=0.48, \,V_{dd}=0.25J$. Similar $T_c$ values are found for the critical temperature of the melting of the PCB phase into a featureless normal fluid, e.g., $T_c\approx 0.35J$ for $V_{dd}=0.25J$. Clearly, for larger interaction strengths, i.e. away from the tip of the lobe, transition temperatures will increase.

{\it Experimental estimates:}
Based on our results  we estimate under which experimental conditions the phases described can be observed. For example, with a gas of Dy ($d=10 \mu_B$) a choice of lattice parameters $a=500$~nm, $d_z=200$~nm, $J=50\, h$Hz results in $V_{dd}/J\sim 0.21$ which stabilizes the PCB phase. In the case of Er$_2$ Feshbach molecules~\cite{FerlainoPrivComm,Dalmonte2010}($d=14 \mu_B$) with $a=400$~nm, $d_z=200$~nm, $J=100\, h$Hz the PCB phase is stabilized at $V_{dd}/J\sim 0.4$. In both cases the PCB phase can be observed at nk temperatures.

Using RbCs ($d=0.3$D) and typical trapping parameters $a=500$~nm, $d_z=300$~nm and $J=150\, h$Hz we find $V_{dd}/J=0.7$, which is large enough to stabilize the PCB. The latter survives up to $T_c^{PCB}\sim 4$~nK. By doping away from filling factor $n=0.5$ the PSS phase can be reached with a KT transition temperature for PSF-normal transition of the order of nK. 

In conclusion, we have studied the quantum phases of dipolar bosons in a bilayer lattice geometry described by the microsopic Hamiltonian Eq.~\eqref{eq:H} for hard-core particles, in a situation where the number of particles  in each layer is the same. Relevant to experiments with polar molecules and magnetic atoms, we have established under which conditions pairing for two particles is stabilized across the layers. Our zero temperature study indicates that the system displays a rich ground state phase diagram including a novel pair-supersolid phase for hard-core dipolar bosons, in addition to pair superfluid and checkerboard-like solid phases. Our finite temperature results indicate that these phases are experimentally observable at temperatures of the order of nK. A four-body correlation function connected with the spread of the pair wave-function can be used to characterize these phases and their transitions. Future work will include the extension of similar quantum Monte Carlo studies to multilayer geometries as well as to systems with population imbalance in the layers.
\\\\
B.C-S. and \c S.G.S. would like to thank A. Kuklov for enlightening discussions. G.P. is supported by EOARD. This work was supported by NSF through a grant to ITAMP at the Harvard-Smithsonian Center for Astrophysics.

\end{document}